\documentstyle[epsfig]{mn}
\newcommand{\Msolar}{\mbox{\,$\rm M_{\odot}$}}        

\begin{document}
\title[black hole--bulge relation in AGNs]{The black hole--bulge relation in AGNs}
\author[W.Bian and Y.Zhao]{W. Bian$^{1,2}$\thanks{E-mail: whbian@njnu.edu.cn} and
Y.Zhao$^{1}$\\
$^{1}$National Astronomical Observatories, Chinese Academy of
 Sciences, Beijing 100012, China\\
$^{2}$Department of Physics, Nanjing Normal University, Nanjing
210097, China}
\date{}
\maketitle
\begin{abstract}
We used the widths of H$\beta$ and [OIII] emission lines to
investigate the black hole--bulge relation in radio-loud AGNs,
radio-quiet AGNs and NLS1s. The central black hole mass, $M_{bh}$,
is estimated from the H$\beta$ line width and the optical
luminosity, and the bulge velocity dispersion, $\sigma$, is
directly from the width of [OIII] line. We found that the
radio-quiet AGNs follow the established $M_{bh}-\sigma$
relationship in nearby inactive galaxies, while the radio-loud
AGNs and NLS1s deviate from this relationship. There are two
plausible interpretations for the deviation of radio-loud AGNs.
One is that the size of broad line regions (BLRs) emitting
H$\beta$ line is overestimated because of the overestimation of
optical luminosity, the other is that the dynamics of BLRs and/or
narrow line regions (NLRs) in radio-loud AGNs is different from
that in radio-quiet AGNs. The deviation of NLS1s may be due to the
small inclination of BLRs to the line of sight or the reliability
of [OIII] line width as the indicator of stellar velocity
dispersion because of its complex multiple components.
\end{abstract}

\section{INTRODUCTION}

Evidence shows that the evolution of black holes and that of their
host galaxies appear to be closely coupled. It was found that
there is strong correlation between the central black hole masses,
$M_{bh}$, and their bulge stellar velocity dispersion, $\sigma$.
Tremaine et al. (2002) investigated this relationship in a sample
of 31 nearby inactive galaxies and gave a better expression as,
\begin{equation}
M_{bh}=10^{8.13}(\sigma/(200 km s^{-1}))^{4.02}\Msolar ,
\end{equation}

There are many methods to estimate the central black hole masses
(Bian \& Zhao 2003a and reference therein). In these methods, the
reverberation method is thought to be more reliable. Using the
reverberation mapping method, the sizes of broad line regions
(BLRs) and then the central black hole masses were obtained for 37
AGNs (Ho 1998; Wandel et al. 1999; Kaspi et al. 2000). For some
AGNs with available bulge velocity dispersion and the
reverberation mapping mass, Gebhardt et al. (2000) and Ferrarese
et al. (2001) also found that these AGNs also follow the $M_{\rm
bh}-\sigma$ relation founded in the nearby inactive galaxies. As
we know, it is difficult to obtain the bulge velocity dispersion
of AGNs. In order to investigate this relation in a larger sample
of AGNs, Nelson (2000) used the width of [OIII] line emitting from
the narrow line region (NLRs) to indicate the bulge velocity
dispersion, where $\sigma=FWHM([OIII])/2.35$, and found that these
37 AGNs with the reverberation mapping masses follow the $M_{\rm
bh}-\sigma$ relation. Wang \& Lu (2001) investigated this relation
in a sample of NLS1s from Veron-Cetty \& Veron (2001). They used
the B band magnitude and the H$\beta$ FWHM to estimate the black
hole masses and the [OIII] FWHM to indicate the bulge velocity
dispersion. They found that NLS1s also follow the $M_{\rm
bh}-\sigma$ relation but with more scatter. We should notice that
NLS1s deviated from the correlation defined in the nearby inactive
galaxies if we think [OIII] FWHM is not overestimated because of
the spectral resolution. Using the Sloan Digital Sky Survey
(SDSS), Boroson (2003) investigated the relation between the black
hole mass via the H$\beta$ FWHM and the stellar velocity
dispersion via [OIII] FWHM in a sample of 107 low-redshift
radio-quiet AGNs. They found the correlation is consistent with
that defined in nearby galaxies and the [OIII] FWHM can predict
black hole mass to a factor of 5. There are only a few radio-loud
AGNs in Boroson (2003). Shields et al. (2003) also investigated
the $M_{\rm bh}-\sigma$ relation as a function of redshift for an
assembled sample of quasars. They suggested that this correlation
can be right out to redshift of $z\approx 3$. However, Shields et
al. (2003) noticed that the radio-loud AGNs seem to deviate from
this correlation.

The central black hole mass can be obtained from the H$\beta$ FWHM
and the optical luminosity, and the bulge velocity dispersion can
be indicated by the [OIII] FWHM. This provides us the opportunity
to investigate the $M_{\rm bh}-\sigma$ relation in a larger sample
of AGNs with available optical spectra. Moreover, we need to
investigate this correlation in a larger sample of radio-loud AGNs
and NLS1s. In next section we present the method to estimate the
black hole mass, and then our adopted data set. Our results and
discussion are given in section 3. Conclusion is presented in the
last section. All of the cosmological calculations in this paper
assume $H_{0}=75 \rm {~km ~s^ {-1}~Mpc^{-1}}$, $\Omega_{M}=0.3$,
$\Omega_{\Lambda} = 0.7$.

\section{METHOD AND DATA}

\subsection{Estimation the Black Hole Masses}

For the reverberation mapping method, it takes a long-term to
simultaneously monitor the variability of the broad emission line
and the continuum, and then to obtain the BLRs size. Up to now,
there are only 37 AGNs with the reverberation mapping mass (Ho
1998; Wandel et al. 1999; Kaspi et al 2000). Fortunately, with the
study of the reverberation mapping method, Kaspi et al. (2000)
found an empirical correlation between the BLRs size and the
monochromatic luminosity at 5100$\rm{\AA}$:
\begin{equation}
\rm{R_{BLR}=32.9(\frac{\lambda L_{\lambda}(5100 \rm{\AA})}{10^{44}
erg \cdot s^{-1}})^{0.7} ~~\rm{lt-days}},
\end{equation}
where $\lambda L_{\lambda}(5100 \rm{\AA)}$ can be estimated from
the optical maginitude by adopting an average optical spectral
index of -0.3 and accounting for Galactic redding and K-correction
(Wang \& Lu 2001). Assuming that the H$\beta$ widths reflect the
Keplerian velocity of the line-emitting BLR material around the
central black hole, we can estimate the viral black hole mass:
\begin{equation}
\rm{M_{\rm bh}=R_{BLR}V^{2}G^{-1}},
\end{equation}
where G is the gravitational constant, V is the velocity of the
line-emitting material. V can be derived from FWHM of the H$\beta$
width. Assuming the random orbits, Kaspi et al.(2000) related $V$
to FWHM of H$\beta$ line by $V=(\sqrt{3}/2) \rm FWHM_{\rm
[H\beta]}$.

This method to estimate the central black hole masses of AGNs has
been discussed by some authors (Wang \& Lu 2001; Bian \& Zhao
2003b; Bian \& Zhao 2003c; Shields et al. 2003; Boroson 2003).

\subsection{Data of NLS1s}

Willams et al. (2003) presented a sample of 150 low-redshift NLS1s
($z<0.8$) found within the SDSS. Using the SDSS Query Tool, we
downloaded the spectra data and the photometry data of these 150
NLS1s. We used the H$\beta$ FWHM from their Table 1. The value of
$\lambda L_{\lambda}(5100 \rm{\AA)}$ is estimated from the
$r^{\ast}$ magnitude. Fluxes were converted to luminosity using
the Schlegel et al.(1998) maps for correcting for Galactic
absorption. We can obtain the central black hole masses in these
150 NLS1s through equation (2) and (3). Each spectrum was shifted
to the rest frame and we preformed a quadratic continuum fit.
Using the splot tools in IRAF software, We measured the [OIII]
FWHM through a Gaussian curve fit to each [OIII] line. The
spectrum resolution R is about 1800, which is equivalent to 166
$km~ s^{-1}$. The error of measured [OIII] FWHM and H$\beta$ FWHM
is about 10\%. The [OIII] FWHM is used to estimate the host
velocity dispersion. It is difficult to measure [OIII] FWHM in
three NLS1s of SDSS J010226.31-003904.6, SDSS J013521.68
-004402.2, and SDSS J15324.367-004342.5 because of their [OIII]
line with irregular profile or low signal-noise ratio, which are
omitted in our discussion. We listed the NLS1s data in Table 1.

\subsection{Data of Radio-Loud and Radio-Quiet AGNs}

For radio-loud and radio-quiet AGNs, we adopted the data of the
widths of H$\beta$ line and [OIII] line from Marziani et al.
(1996). Marziani et al. (1996) used a sample of 52 low-redshift
($z<0.8$) AGNs with available UV and optical spectra to do a
comparative analysis of high-ionization and low-ionization lines
in BLRs. There are 31 radio-loud AGNs and 21 radio-quiet AGNs in
their sample. They found radio-loud and radio-quiet AGNs show
strong difference on the dynamic of emission lines in BLRs. The
sample is suitable to study the difference if any on
$M_{bh}-\sigma$ relation between radio-loud and radio-quiet AGNs.
For FWHM of broad component of H$\beta$ line, FeII emission and
the narrow component were subtracted. We obtained FWHM of H$\beta$
line from Column (10) and (12) in Table 8 from Marziani et al.
(1996). The absolute optical B band magnitude for these 52 AGNs
are adopted from Veron-Cetty \& Veron (2001). The [OIII] line
widths are adopted from Column (17) in Table 5 from Marziani et
al. (1996). There are 48 AGNs with available widths of H$\beta$
and [OIII] line. Using equation (2) and (3), we obtained the
central black hole masses of 48 AGNs in the sample of Marziani et
al. (1996), including 12 flat-spectrum AGNs, 18 steep-spectrum
AGNs, 18 radio-quiet AGNs. The date are listed in Table 2.

\section{RESULTS AND DISCUSSION}

\begin{figure}
\centerline{\epsfxsize=100mm\epsfbox{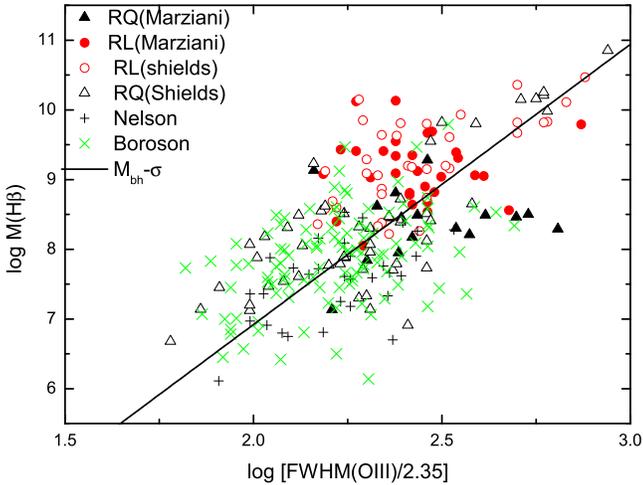}} \caption{Black hole
masses derived from H$\beta$ line width and B band magnitude
versus width of the [OIII] line for AGNs. The solid line shows the
$M_{\rm bh}-\sigma$ relation from equation (1). Open circle:
radio-quiet AGNs in Shields et al. (2003); solid circle:
radio-loud AGNs in Shields et al. (2003) ; solid triangle:
radio-loud AGNs from Marziani et al. (1996); open triangle:
radio-quiet AGNs from Marziani et al. (1996); plus: AGNs from
Nelson (2001); fork: AGNs from Boroson (2003).}
\end{figure}

\begin{figure}
\centerline{\epsfxsize=100mm\epsfbox{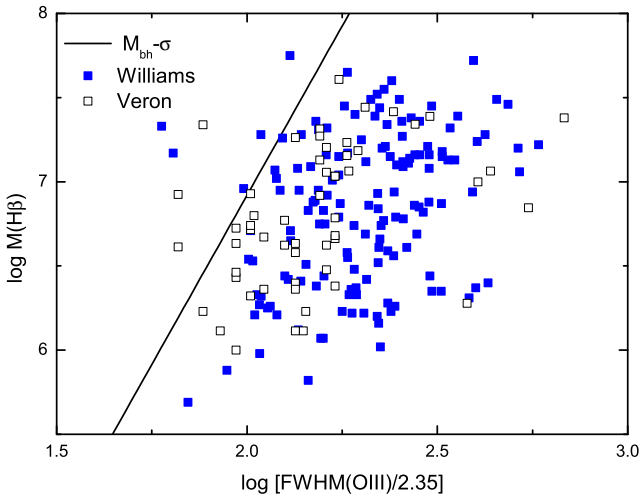}} \caption{Black hole
masses derived from H$\beta$ line width and B band magnitude
versus width of the [OIII] line for Narrow-line AGNs. The solid
line shows the $M_{\rm bh}-\sigma$ relation from equation (1).
Open square: NLS1S from Wang \& Lu (2001); solid square: NLS1s
from Williams et al. (2003).}
\end{figure}

\subsection{$M_{\rm H\beta}$--$\sigma_{\rm [OIII]}$ Relation}

In Fig. 1 and Fig. 2, we plot black hole masses estimated from the
H$\beta$ line width versus the bulge velocity dispersion obtained
from the [OIII] line width for the samples of Marziani et al.
(1996), Shields et al. (2003), Wang \& Lu (2001), Nelson (2001),
Boroson (2003), and Williams et al. (2003).

The sample of Shields et al. (2003) included 49 radio-quiet AGNs
and 35 radio-loud AGNs. Shields et al. (2003) used the H$\beta$
emission line width to investigate the $M_{\rm bh}-\sigma$
relation as a function of redshift for an assembled sample of
quasars. They suggested that this correlation is not a function of
redshift and can be right out to redshift of $z\approx 3$. They
adopted the relation between BLRs sizes and continuum luminosity
suggested by photoionization model, $R \propto L^{0.5}$. In order
to consistent with our calculation, we recalculated the black hole
masses of AGNs in their sample using the equation (2) and (3). The
sample of Wang \& Lu (2001) consisted of 59 NLS1s from Veron-Cetty
et al. (2001). The sample of Williams et al. (2003) included 147
NLS1s from SDSS. Up to now, this is the largest sample of NLS1s.

From Fig. 1 and Fig. 2, it is found that radio-quiet AGNs in
Marziani et al. (1996), Shields et al. (2003), and Boroson (2003)
follow the $M_{\rm bh}$--$\sigma$ relation defined in equation (1)
with a larger scatter compared with that in Nelson (2001).
However, the radio-loud AGNs in Marziani et al. (1996) and Shields
et al. (2003), and NLS1s in Wang \& Lu (2001) and Williams et al.
(2003) seemed not follow this relation.

For radio-loud AGNs, the mean black hole mass estimated from
H$\beta$ FWHM is larger than that from [OIII] FWHM. For NLS1s, the
mean black hole mass estimated from H$\beta$ FWHM is smaller than
that from [OIII] FWHM. In Fig. 1 and Fig. 2, it is clear that
radio-loud AGNS and NLS1s deviated from the relation defined in
equation (1).

We also calculated the black hole mass, $M_{[\rm OIII]}$, directly
from the equation (1) using [OIII] line width as the indicator of
$\sigma$. Table 3 showed the distribution of log($M_{\rm
H\beta}/M_{[\rm OIII]})$ for different AGNs samples. The
distribution of log($M_{\rm H\beta}/M_{[\rm OIII]})$ for
radio-quiet AGNs in Marziani et al. (1996) is $-0.36\pm0.19$ with
a standard deviation of 0.81 (See Table 3). It showed that the
mass estimated from H$\beta$ line width is consistent with that
from the [OIII] line width but with a large scatter, which is
consistent with the data of radio-quiet AGNs in Fig.1. However,
the distribution of log($M_{\rm H\beta}/M_{[\rm OIII]})$ for
radio-loud AGNs in Marziani et al. (2003) is $0.51\pm0.13$ with a
standard deviation of 0.73. It showed that the mass estimated from
H$\beta$ line width is larger than that from the [OIII] line
width, which is also consistent with the data of radio-loud AGNs
in Fig.1. The distributions of log($M_{\rm H\beta}/M_{[\rm
OIII]})$ for radio-loud and radio-quiet AGNs in Shields et al.
(2003) are listed in Table 3. The results for the sample of
Shields et al. (2003) are consistent with that for the sample of
Marziani et al. (1996).

\subsection{Uncertainties of Black Hole Mass and Stellar Velocity Dispersion}

In our analysis, we used equations (2)-(3) to calculate the black
hole masses and FWHM of [OIII] line to indicate the bulge velocity
dispersion. The errors of the calculated black hole masses using
equations (2)-(3) are mainly from the accuracy of equation
(2)-(3); the geometry and the dynamics of the BLRs, especially the
disk inclination to the line of sight in NLS1s (Bian \& Zhao 2002)
and in flat-spectrum quasars (Jarvis \& McLure 2003). The
appropriate measurement of H$\beta$ line width for estimating the
black hole mass were discussed by some authors (Vestergaard 2002;
Shields et al. 2003). The error in the mass estimation using
equations (2) and (3) is about 0.5 dex (Wang \& Lu 2001).

It is possible to measure the luminosity at 5100$\AA$ ($L_{spec}$)
for 147 SDSS NLS1s spectrum. We compared the luminosity at
5100$\AA$ estimated from $r^{*}$ ($L_{r}$) with that from SDSS
spectrum in Fig. 3. They are consistent and the distribution of
log($L_{r}/L_{spec}$) is 0.21$\pm$0.01 with a standard deviation
of 0.15. The mean mass estimation using $L_{r}$ would be enhanced
by 0.15 dex compared with that using $L_{spec}$.

McLure \& Dunlop (2001) suggested that the assumption of random
orbits of BLRs seems unrealistic for quasars, and that the BLRs
velocity should be related to FWHM of H$\beta$ as $V=1.5\times \rm
FWHM_{\rm [H\beta]}$. It is equivalent to assume smaller
inclination in quasars (see a detail discussion in McLure \&
Dunlop 2001). Gu et al. (2001) also used $V=1.5\times \rm
FWHM_{\rm [H\beta]}$ to estimate the velocity of BLRs in
radio-loud AGNs. Jarvis \& McLure (2002) investigated the relation
between the black hole mass and radio luminosity in flat-spectrum
quasars. Considering the Doppler boosting correction of the radio
luminosity and the inclination correction of the BLRs velocity,
they found flat-spectrum quasars follow the relation between radio
luminosity and black hole mass found by Dunlop et al. (2003).
Jarvis \& McLure (2002) adopted a correction factor of two for the
BLRs velocity of flat-spectrum quasars as the effect of
inclination, which would increase the black hole mass estimates by
a factor of four, $\sim 0.6 (dex)$. Smaller inclination in
radio-loud AGNs would enhance the value of BLRs virial velocity
derived from H$\beta$ line width and would make radio-loud AGNs
deviate much from the line defined by equation (1) in Fig.1.

The [OIII] line width may be overestimated by a factor of 1.3
because of the poor resolution spectrum (Veilleux 1991; Wang \& Lu
2001). Considering the overestimation of the width of [OIII] line,
we found that NLS1s in Wang \& Lu (2001) and SDSS NLS1s are also
deviated from the relation defined in equation (1) (see Fig. 2).
The overestimation of bulge velocity dispersion derived from the
observed [OIII] line width would also make radio-loud AGNs deviate
much from the line defined in equation (1).

Radio-loud AGNs with significant H$\beta$ blueshifts or redshifts
have been observed (Marziani et al. 1996). The oversimple models
involving pure rotation or radial motion are unlikely. There are
many possible components of motion in the BLRs (Marzinai et al.
1996). For luminous AGNs, bright emission lines would contribute
to the optical continuum luminosity, which would overestimate the
BLRs sizes derived from equation (2). It's the use of a broad band
luminosity (optical magnitude) not being converted to the
luminosity in 5100 \AA properly that also accounts for the
overestimate of BLRs sizes (also see Fig. 3). The optical
luminosity may be contaminated by the synchrotron emission from
the jet for flat-spectrum quasars whose radio emission is beamed
to us (Gu et al. 2001, Jarvis \& McLure 2002). The overestimated
optical continuum luminosity would overestimate BLRs sizes, which
would account for the overestimated black hole masses in
radio-loud AGNs. This would lead to the deviation from the
relation defined in equation (1) for radio-loud AGNs. If we think
all type of galaxies follow the same $M_{\rm bh}-\sigma$
correlation, we should be cautious to use equation (2)-(3) to
calculate the black hole masses for radio-loud AGNs.

The overestimation of [OIII] FWHM from poor resolution spectrum in
NLS1s is about a factor of 1.3, which would lead to about 0.4 dex
in black hole mass. For NLS1s, we should consider other causes for
the deviation from the correlation defined in equation (1). There
are mainly several opinions about the origin about the narrow
width of H$\beta$ in NLS1s. One is the small inclinations in NLS1s
(Fig.1 in McLure \& Dunlop 2002; Bian \& Zhao 2002); the second is
the long distance of BLRs emitting line of H$\beta$ in NLS1s; the
third is their higher value of $L/L_{Edd}$ because of their low
central black hole masses. From Fig. 2, the smaller inclination in
NLS1s is possible if we think NLS1s follow the correlation defined
by equation (1). Nelson \& Whittle (1996) showed in their Fig. 8
that the velocity dispersion from the [OIII] FWHM of the more
radio luminous objects tend to be overestimated, which would make
radio-loud AGNs deviate much from the correlation defined in
equation (1). Tadhunter et al. (2001) found that [OIII] of radio
galaxy PKS 1549-79 is unusually broad and is blueshifted related
to the low-ionization lines. They suggested there is an outflow in
an inner NLRs. Recently Holt et al. (2003) investigated the
intermediate resolution spectra ($\sim$ 4$\AA$) of a radio source
PKS 1345+12 and found there are three complex components on [OIII]
emission line, which would affect the measurement of narrow [OIII]
FWHM as a tracer of host velocity dispersion. For radio-loud AGNs,
the [OIII] profile would be complex because of the interaction
between radio jet and the interstellar medium (Gelderman et al.
1994). The research on the relation between narrow component of
[OIII] line and the host velocity dispersion, especially for
radio-loud AGNs, is needed.

\begin{figure}
\centerline{\epsfxsize=100mm\epsfbox{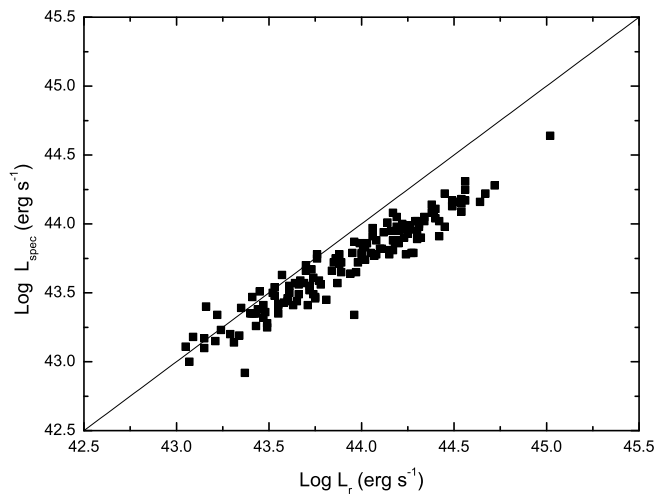}}
\caption{Monochromatic luminosity at 5100$\AA$ measured from the
spectra of the SDSS NLS1s versus that from $r^{*}$ band
magnitude.}
\end{figure}

\section{Conclusion}
The correlation between the central black hole mass and the bulge
velocity dispersion was investigated in radio-quiet AGNs,
radio-loud AGNs and NLS1s. The main conclusions can be summarized
as follows:
\begin{itemize}
\item{The radio-quiet AGNs follow the $M_{\rm bh}-\sigma$ correlation
defined by equation (1) founded in nearby inactive galaxies, while
radio-loud AGNs and NLS1s seem not follow this correlation.}
\item{Small inclination or overestimated bulge velocity
dispersion can not account for the deviation of radio-loud AGNs
from the correlation defined in equation (1). There are two
possibility to explain this deviation in radio-loud AGNs. One is
that the size of BLRs emitting H$\beta$ line is overestimated
because of the overestimation of optical luminosity, the other is
that the dynamics in BLRs and/or NLRs in radio-loud AGNs is
different from that in radio-quiet AGNs.}
\item{For NLS1s, small inclination may play a particular role in
the deviation from the correlation defined in equation (1). We
should consider the inclination effect in the black hole mass
estimation using H$\beta$ FWHM. }

\end{itemize}

\section*{ACKNOWLEDGMENTS}
We thank the anonymous referee for the valuable comments. This
work has been supported by the NSFC (No. 10273007) and NSF of
Jiangsu Provincial Education Department (No. 03KJB160060). The
SDSS web site is http://www.sdss.org/. This research has made use
of the NASA/IPAC Extragalactic Laboratory Database (NED), which is
operated by the Jet Propulsion Laboratory, California Institute of
Technology, under contract to NASA.

\begin{table}
\begin{center}
\begin{small}
\begin{tabular}{lcccccccc}
\hline\hline
name&z&H$\beta$&$\nu L_{\nu}$&$M_{\rm bh}$&$\sigma_{[\rm OIII]}$\\
(1)& (2)& (3)&(4)&(5)&(6)\\
\hline
J000109.14-004121.5 &0.417&1209 & 44.11 & 6.92 &   2.21\\
J000834.72+003156.2 &0.263&1351 & 44.38 & 7.21 &   2.36\\
J001327.31+005232.0 &0.363&1742 & 44.56 & 7.55 &   2.36\\
J002213.00-004832.7 &0.214&1429 & 43.35 & 6.54 &   2.00\\
J002233.27-003448.6 &0.504&1388 & 44.26 & 7.15 &   2.38\\
J002305.03-010743.5 &0.166&1157 & 43.57 & 6.51 &   2.15\\
J002752.39+002615.8 &0.205&1830 & 43.87 & 7.12 &   2.31\\
J003024.94+000254.5 &0.288&743  & 43.72 & 6.23 &   2.25\\
J003238.20-010035.2 &0.092&639  & 43.40 & 5.88 &   1.95\\
J003431.74-001312.7 &0.381&1314 & 44.42 & 7.21 &   2.47\\
J003711.00+002128.0 &0.235&617  & 43.85 & 6.16 &   2.35\\
J004052.14+000057.3 &0.405&1278 & 44.49 & 7.24 &   2.60\\
J004338.54-005814.7 &0.559&1122 & 44.45 & 7.10 &   2.39\\
J005446.16+004204.1 &0.234&1225 & 43.78 & 6.71 &   2.11\\
J005921.37+004108.9 &0.423&1625 & 44.08 & 7.16 &   2.48\\
J011357.93-011139.8 &0.754&1842 & 44.72 & 7.72 &   2.60\\
J011703.58+000027.4 &0.046&975  & 43.16 & 6.07 &   2.19\\
J011712.81-005817.5 &0.486&1937 & 44.20 & 7.40 &   2.28\\
J011929.06-000839.7 &0.090&900  & 43.44 & 6.20 &   2.34\\
J013046.16-000800.8 &0.253&1648 & 43.63 & 6.86 &   2.44\\
J013521.68-004402.2 &0.098&1181 & 43.45 & 6.44 &   2.48\\
J013842.05+004020.0 &0.520&1035 & 44.24 & 6.88 &   2.48\\
J013940.99-010944.4 &0.194&1091 & 43.71 & 6.55 &   2.26\\
J014234.41-011417.4 &0.244&1607 & 43.96 & 7.07 &   2.07\\
J014412.77-000610.5 &0.359&1041 & 43.66 & 6.48 &   2.28\\
J014542.78+005314.9 &0.389&1255 & 43.86 & 6.78 &   2.41\\
J014559.45+003524.7 &0.166&1075 & 43.43 & 6.35 &   2.49\\
J014644.82-004043.2 &0.083&1164 & 43.22 & 6.27 &   2.03\\
J014951.66+002536.5 &0.252&563  & 43.70 & 5.98 &   2.03\\
J015652.43-001222.0 &0.163&1324 & 43.67 & 6.69 &   2.31\\
J020431.64+002400.5 &0.171&1077 & 43.29 & 6.25 &   2.05\\
J021610.56+000538.4 &0.384&1467 & 44.17 & 7.13 &   2.53\\
J021652.47-002335.3 &0.304&854  & 44.06 & 6.59 &   2.37\\
J022205.37-004948.0 &0.525&1571 & 44.54 & 7.45 &   2.26\\
J022756.28+005733.1 &0.128&773  & 43.09 & 5.82 &   2.16\\
J022841.48+005208.6 &0.186&990  & 43.64 & 6.42 &   2.31\\
J022923.43-000047.9 &0.558&1386 & 44.40 & 7.25 &   2.30\\
J023057.39-010033.7 &0.649&1947 & 44.56 & 7.65 &   2.26\\
J023211.83+000802.4 &0.432&1746 & 44.00 & 7.16 &   2.45\\
J023414.58+005707.9 &0.269&1381 & 43.75 & 6.79 &   2.24\\
J024037.89+001118.9 &0.470&1789 & 44.19 & 7.32 &   2.21\\
J024651.91-005931.0 &0.468&1504 & 45.02 & 7.75 &   2.11\\
J025501.19+001745.5 &0.360&904  & 43.77 & 6.44 &   2.20\\
J030031.31+005357.2 &0.198&1536 & 43.55 & 6.74 &   2.28\\
J030417.78+002827.4 &0.044&1321 & 43.05 & 6.26 &   2.06\\
J030639.57+000343.2 &0.107&1525 & 43.70 & 6.84 &   2.34\\
J031427.47-011152.4 &0.387&1812 & 44.45 & 7.52 &   2.34\\
J031542.64+001228.7 &0.207&870  & 43.73 & 6.37 &   2.60\\
J031630.79-010303.6 &0.368&1226 & 44.12 & 6.94 &   2.39\\
J032255.49+001859.9 &0.384&1621 & 44.49 & 7.45 &   2.48\\
J032337.65+003555.7 &0.215&1490 & 44.15 & 7.13 &   2.55\\
J032606.75+011429.9 &0.127&686  & 43.52 & 6.02 &   2.35\\
J033027.21+005433.7 &0.443&1315 & 44.34 & 7.16 &   2.21\\
J033059.06+010952.1 &0.557&1946 & 44.15 & 7.36 &   2.41\\
J033429.44+000611.0 &0.347&1316 & 44.67 & 7.39 &   2.34\\
J033854.25+005339.7 &0.279&1314 & 43.81 & 6.79 &   2.39\\
J033923.66-002310.3 &0.369&1437 & 43.94 & 6.95 &   2.14\\
J034131.95-000933.0 &0.223&897  & 43.53 & 6.26 &   2.39\\
J034326.51+003915.2 &0.499&1315 & 44.32 & 7.15 &   2.24\\
J034430.03-005842.7 &0.287&786  & 43.89 & 6.40 &   2.63\\
J094857.33+002225.5 &0.584&1342 & 44.56 & 7.33 &   1.78\\
J095859.80+004718.9 &0.235&1190 & 43.75 & 6.66 &   2.35\\
J100405.00-003253.4 &0.289&582  & 44.14 & 6.31 &   2.58\\

\hline
\end{tabular}
\caption{$M_{\rm bh}$ and $\sigma_{[\rm OIII]}$ for NLS1s in SDSS.
Col. (1): Object name. Col. (2): Redshift. Col. (3):FWHM of the
broad H$\beta$ line in units of $\rm km~ s^{-1}$.  Col. (4): log
of continuum luminosity at 5100$\AA$ rest wavelength in units of
$\rm erg~ s^{-1}$. Col. (5): log of the black hole mass in units
of solar mass. Col. (6): log of the bulge velocity dispersion
derived from FWHM of [OIII] line in units of $\rm erg~ s^{-1}$.}
\end{small}
\end{center}
\end{table}

\begin{table}
\begin{center}
\begin{small}
\begin{tabular}{lcccccccc}
\hline\hline
J101314.86-005233.5 & 0.276&1578 & 44.29 & 7.28   &   2.63\\
J102059.72+010034.3 & 0.588&1715 & 44.64 & 7.60   &   2.38\\
J102450.52-002102.4 & 0.322&1382 & 44.17 & 7.08   &   2.13\\
J103031.41-001902.6 & 0.562&1787 & 44.22 & 7.34   &   2.37\\
J103222.58-000345.6 & 0.559&1707 & 44.19 & 7.28   &   2.04\\
J103457.29-010209.0 & 0.328&1394 & 44.49 & 7.31   &   2.19\\
J104132.35-003512.2 & 0.135&1316 & 43.15 & 6.33   &   2.03\\
J104210.03-001814.7 & 0.115&628  & 43.15 & 5.69   &   1.85\\
J104230.14+010223.7 & 0.116&1012 & 43.41 & 6.28   &   2.37\\
J104331.51-010732.9 & 0.362&1756 & 44.05 & 7.20   &   2.35\\
J104449.28+000301.2 & 0.443&1176 & 44.07 & 6.87   &   2.51\\
J105932.52-004354.7 & 0.155&1451 & 43.37 & 6.56   &   2.39\\
J110312.83+000012.5 & 0.276&1450 & 43.89 & 6.93   &   2.34\\
J111022.39-005544.5 & 0.257&1934 & 43.88 & 7.17   &   2.26\\
J111307.73+003210.4 & 0.346&976  & 44.16 & 6.77   &   2.36\\
J113102.28-010122.0 & 0.242&1928 & 43.49 & 6.89   &   2.18\\
J113541.20+002235.4 & 0.175&1165 & 44.03 & 6.83   &   2.16\\
J115023.59+000839.1 & 0.127&1136 & 43.49 & 6.44   &   2.10\\
J115306.95-004512.7 & 0.357&1102 & 43.98 & 6.75   &   2.20\\
J115412.77+010133.4 & 0.490&945  & 44.31 & 6.85   &   2.45\\
J115533.50+010730.6 & 0.197&1628 & 43.76 & 6.94   &   2.59\\
J115755.47+001704.0 & 0.261&1762 & 43.97 & 7.16   &   2.44\\
J115832.81+005139.2 & 0.591&1035 & 44.54 & 7.09   &   2.17\\
J121415.17+005511.4 & 0.396&1981 & 44.30 & 7.49   &   2.32\\
J122102.95-000733.7 & 0.366&517  & 44.17 & 6.23   &   2.38\\
J122432.40-002731.4 & 0.157&1308 & 43.76 & 6.75   &   2.19\\
J124519.73-005230.4 & 0.221&1730 & 43.53 & 6.83   &   2.20\\
J125337.36-004809.6 & 0.427&1416 & 44.31 & 7.20   &   2.71\\
J125943.59+010255.1 & 0.394&1459 & 44.30 & 7.22   &   2.77\\
J130023.22-005429.8 & 0.122&1018 & 43.58 & 6.41   &   2.14\\
J130707.71-002542.9 & 0.450&1475 & 44.18 & 7.15   &   2.51\\
J130855.18+004504.1 & 0.429&1851 & 44.06 & 7.26   &   2.09\\
J131108.48+003151.8 & 0.429&1642 & 44.54 & 7.49   &   2.66\\
J132231.13-001124.5 & 0.173&1861 & 43.61 & 6.95   &   2.19\\
J133031.41-002818.8 & 0.240&1216 & 43.60 & 6.58   &   2.26\\
J133741.76-005548.2 & 0.279&873  & 43.69 & 6.35   &   2.51\\
J135908.01+002732.0 & 0.257&1282 & 43.95 & 6.87   &   2.24\\
J141234.68-003500.0 & 0.127&1098 & 43.21 & 6.21   &   2.08\\
J141519.50-003021.6 & 0.135&1186 & 43.34 & 6.37   &   2.29\\
J141820.33-005953.7 & 0.254&831  & 43.72 & 6.33   &   2.27\\
J142441.21-000727.1 & 0.318&1201 & 44.23 & 7.01   &   2.22\\
J143030.22-001115.1 & 0.103&1744 & 43.07 & 6.52   &   2.34\\
J143230.99-005228.9 & 0.362&1559 & 43.99 & 7.06   &   2.72\\
J143624.82-002905.3 & 0.325&1857 & 44.34 & 7.46   &   2.69\\
J144735.25-003230.5 & 0.217&1105 & 43.65 & 6.53   &   2.01\\
J144913.51+002406.9 & 0.441&944  & 44.08 & 6.69   &   2.45\\
J144932.70+002236.3 & 0.081&1072 & 43.24 & 6.21   &   2.02\\
J145123.02-000625.9 & 0.139&1122 & 43.43 & 6.38   &   2.18\\
J145437.84-003706.6 & 0.576&1328 & 44.28 & 7.12   &   2.41\\
J150629.23+003543.2 & 0.370&1861 & 44.19 & 7.36   &   2.18\\
J151312.41+001937.5 & 0.159&1697 & 43.60 & 6.86   &   2.32\\
J151956.57+001614.6 & 0.115&1716 & 43.61 & 6.88   &   2.17\\
J153911.17+002600.8 & 0.265&539  & 44.10 & 6.22   &   2.31\\
J164907.64+642422.3 & 0.184&759  & 43.47 & 6.07   &   2.20\\
J165022.88+642136.1 & 0.407&1152 & 44.02 & 6.82   &   2.46\\
J165338.69+634010.7 & 0.279&1848 & 44.25 & 7.39   &   2.55\\
J165537.78+624739.0 & 0.597&1271 & 44.40 & 7.17   &   1.81\\
J165633.87+641043.7 & 0.272&1139 & 43.74 & 6.61   &   2.42\\
J165658.38+630051.1 & 0.169&1466 & 43.47 & 6.65   &   2.11\\
J165905.45+633923.6 & 0.368&1359 & 44.20 & 7.09   &   2.41\\
J170546.91+631059.1 & 0.119&1657 & 43.41 & 6.71   &   2.01\\
J170812.29+601512.6 & 0.145&1094 & 43.42 & 6.36   &   2.27\\
J170956.02+573225.5 & 0.522&1329 & 44.50 & 7.28   &   2.14\\
J171033.21+584456.8 & 0.281&652  & 43.88 & 6.22   &   2.28\\
J171207.44+584754.5 & 0.269&1708 & 44.18 & 7.27   &   2.41\\
J171540.92+560655.0 & 0.297&1752 & 44.01 & 7.18   &   2.51\\
J171829.01+573422.4 & 0.101&1322 & 43.55 & 6.61   &   2.35\\
\hline
\end{tabular}
\end{small}
\end{center}
\end{table}

\begin{table}
\begin{center}
\begin{small}
\begin{tabular}{lcccccccc}
\hline\hline
J172007.96+561710.7& 0.389& 1221 & 43.84 & 6.74 &   2.35\\
J172206.04+565451.6& 0.426& 1579 & 44.39 & 7.36 &   2.45\\
J172756.86+581206.0& 0.414& 1742 & 43.96 & 7.14 &   2.43\\
J172800.67+545302.8& 0.246& 1583 & 44.00 & 7.08 &   2.48\\
J172823.61+630933.9& 0.439& 1750 & 44.30 & 7.38 &   2.43\\
J173404.85+542355.1& 0.685& 1163 & 44.42 & 7.11 &   2.42\\
J173721.14+550321.7& 0.333& 1256 & 44.22 & 7.04 &   2.24\\
J232525.53+001136.9& 0.491& 1921 & 44.27 & 7.44 &   2.35\\
J233032.95+000026.4& 0.123& 956  & 43.55 & 6.33 &   2.29\\
J233149.49+000719.5& 0.367& 1708 & 44.25 & 7.32 &   2.53\\
J233853.83+004812.4& 0.170& 1011 & 43.48 & 6.32 &   2.04\\
J234050.53+010635.6& 0.358& 729  & 44.02 & 6.42 &   2.11\\
J234141.50-003806.7& 0.319& 1871 & 44.38 & 7.49 &   2.40\\
J234150.81-004329.2& 0.251& 1817 & 43.66 & 6.96 &   1.99\\
J234216.74+000224.1& 0.185& 917  & 43.31 & 6.12 &   2.13\\
J234229.46-004731.6& 0.316& 1857 & 43.72 & 7.02 &   2.08\\
J234725.30-010643.7& 0.182& 1667 & 43.74 & 6.95 &   2.09\\
\hline
\end{tabular}

\end{small}
\end{center}
\end{table}

\begin{table}
\begin{center}
\begin{tabular}{lccccccc}
\hline\hline
name&RL/RQ&z&H$\beta$&$\nu L_{\nu}$&$M_{\rm bh}$&$\sigma_{[\rm OIII]}$\\
(1)& (2)& (3)&(4)&(5)&(6)&(7)\\
\hline
0044+030      &   SS &0.6232 & 5480   &   45.75  &    9.13   &   2.54  \\
0050+124      &   RQ &0.0604 & 4460   &   44.47  &    8.29   &   2.81  \\
0121-590      &   RQ &0.0460 & 6080   &   44.31  &    8.45   &   2.39  \\
0349-146      &   SS &0.6162 & 9470   &   45.67  &    9.55   &   2.38  \\
0403-132      &   FS &0.5705 & 4490   &   45.27  &    8.64   &   2.42  \\
0405-123      &   FS &0.5725 & 4830   &   46.19  &    9.35   &   2.42  \\
0414-060      &   SS &0.7750 &14820   &   46.03  &   10.13   &   2.38  \\
0454-220      &   SS &0.5334 & 6920   &   45.67  &    9.31   &   2.54  \\
0710+118      &   SS &0.7700 &19530   &   45.67  &   10.12   &   2.27  \\
0742+318      &   FS &0.4610 & 7240   &   45.75  &    9.43   &   2.23  \\
0838+133      &   SS &0.6808 & 3090   &   44.95  &    8.05   &   2.29  \\
0850+440      &   RQ &0.6139 & 3080   &   45.55  &    8.49   &   2.62  \\
0918+511      &   RQ &0.5563 & 4540   &   44.71  &    8.26   &   2.73  \\
0923+392      &   FS &0.6948 & 8210   &   45.23  &    9.09   &   2.38  \\
0953+414      &   RQ &0.2341 & 3010   &   45.35  &    8.49   &   2.44  \\
0955+326      &   SS &0.5305 & 4730   &   45.79  &    9.06   &   2.59  \\
1001+292      &   RQ &0.3297 & 2450   &   45.27  &    8.21   &   2.57  \\
1007+417      &   SS &0.6123 & 2860   &   45.91  &    8.68   &   2.46  \\
1049-005      &   RQ &0.3599 & 4510   &   45.39  &    8.81   &   2.38  \\
1100+772      &   SS &0.3115 & 7840   &   45.43  &    9.34   &   2.38  \\
1103-006      &   SS &0.4233 & 6600   &   45.39  &    9.12   &   2.44  \\
1116+215      &   RQ &0.1765 & 3130   &   45.23  &    8.46   &   2.70  \\
1136-374      &   RQ &0.0096 & 3780   &   42.99  &    7.13   &   2.21  \\
1137+660      &   SS &0.6460 & 4950   &   45.75  &    9.03   &   2.31  \\
1202+281      &   RQ &0.1653 & 5020   &   44.87  &    8.62   &   2.33  \\
1211+143      &   RQ &0.0809 & 2320   &   44.71  &    7.88   &   2.24  \\
1216+069      &   RQ &0.3313 & 5790   &   45.51  &    9.13   &   2.16  \\
1226+023      &   FS &0.1575 & 3650   &   45.87  &    9.05   &   2.61  \\
1253-055      &   FS &0.5362 &21610   &   44.95  &    9.79   &   2.87  \\
1302-102      &   FS &0.2784 & 3850   &   45.55  &    8.82   &   2.48  \\
1351+640      &   RQ &0.0882 & 3670   &   44.75  &    8.30   &   2.54  \\
1411+442      &   RQ &0.0896 & 2600   &   44.99  &    8.17   &   2.42  \\
1415+253      &   RQ &0.0167 & 6200   &   43.39  &    7.84   &   2.30  \\
1444+407      &   RQ &0.2673 & 2840   &   45.39  &    8.45   &   2.36  \\
1512+370      &   SS &0.3707 & 9360   &   45.35  &    9.41   &   2.34  \\
1538+477      &   RQ &0.7721 & 4920   &   46.19  &    9.28   &   2.46  \\
1545+210      &   SS &0.2643 & 6730   &   44.83  &    8.81   &   2.41  \\
1618+177      &   SS &0.5551 &11530   &   45.59  &    9.69   &   2.47  \\
1637+574      &   FS &0.7506 & 4620   &   45.71  &    8.90   &   2.46  \\
1641+399      &   FS &0.5928 & 4870   &   45.75  &    9.04   &   2.50  \\
1704+608      &   SS &0.3721 & 6990   &   45.71  &    9.41   &   2.27  \\
1928+738      &   FS &0.3021 & 3120   &   45.23  &    8.40   &   2.22  \\
2041-109      &   RQ &0.0344 & 3090   &   44.43  &    7.95   &   2.38  \\
2135-147      &   SS &0.2003 & 7570   &   45.03  &    9.08   &   2.19  \\
2141+175      &   FS &0.2111 & 4320   &   44.99  &    8.56   &   2.68  \\
2201+315      &   FS &0.2950 & 3410   &   45.31  &    8.54   &   2.46  \\
2251+113      &   SS &0.3255 & 4540   &   45.31  &    8.78   &   2.41  \\
2308+098      &   SS &0.4333 &11330   &   45.51  &    9.67   &   2.46  \\
\hline
\end{tabular}
\caption{$M_{\rm bh}$ and $\sigma_{[\rm OIII]}$ for AGNs. Col.
(1): Object name. Col. (2):RQ denotes Radio quiet objects, SS
denotes steep-spectrum objects, and FS denotes flat-spectrum
objects. Col. (3): Redshift. Col. (4): FWHM of the broad H$\beta$
line in units of $\rm km~ s^{-1}$. Col. (5): log of continuum
luminosity at 5100$\AA$ rest wavelength in units of $\rm erg~
s^{-1}$. Col. (6): log of the black hole mass in units of solar
mass. Col. (7): log of the bulge velocity dispersion derived from
FWHM of [OIII] line in units of $\rm m~ s^{-1}$.}
\end{center}
\end{table}

\begin{table}
\begin{center}
\begin{tabular}{llllll}
\hline\hline
Type & log($M_{\rm H\beta}/M_{[\rm OIII]}$)& SD \\
\hline
RL(Marziani)& 0.51$\pm$ 0.13 & 0.73 \\
FS(Marziani)& 0.13$\pm$0.2 & 0.70\\
SS(Marziani)& 0.74$\pm$0.16 &0.66 \\
RL(Shileds) & 0.59$\pm$0.10 & 0.61\\
RQ(Marziani)& -0.36$\pm$0.19 & 0.81\\
RQ(Shields) & 0.17$\pm$0.10 & 0.68\\
RQ(Boroson) & 0.08$\pm$0.07 & 0.75\\
NLS1s(Wang) & -0.84$\pm$0.11& 0.79 \\
NLS1s(Williams) & -1.29$\pm$0.06 & 0.75\\

\hline \hline
\end{tabular}
\caption{The distributions of log($M_{\rm H\beta}/M_{[\rm OIII]}$)
for different type of AGNs. RL(Marziani): Radio-loud AGNs in
Marziani et al. (1996); FS(Marziani): Flat-spectrum AGNs in
Marziani et al. (1996);  SS(Marziani): Steep-spectrum AGNs in
Marziani et al. (1996); RQ(Marziani): Radio-quiet AGNs in Marziani
et al. (1996); RL(Shields): Radio-loud AGNs in Shields et al.
(2003); RQ(Shields): Radio-quiet AGNs in Shields et al. (2003); RQ
AGNs(Boroson): Radio-quiet AGNs in Boroson (2003); NLS1s(Wang):
NLS1s in Wang \& Lu (2001); NLS1s(Williams): NLS1s in Williams et
al.(2003). }
\end{center}
\end{table}

\end{document}